\documentclass[conference]{IEEEtran}

\usepackage{amsmath,amsfonts,amssymb,dsfont,subfigure,amssymb,vaucanson-g,graphicx,color}
\usepackage{authblk}
\usepackage{vaucanson-g}
\usepackage{tikz}
\usetikzlibrary{arrows}

\newcommand{\rea}{\mathbb{R}}

\newtheorem{theorem}{\textbf{Theorem}}

\newtheorem{lemma}[theorem]{\textbf{Lemma}}

\IEEEoverridecommandlockouts

\begin{document}

\title{Consensus in multi-agent systems \\ with non-periodic sampled-data exchange \\ and uncertain network topology}

\author{Mehran Zareh$^*$, Dimos V. Dimarogonas$^{**}$, Mauro Franceschelli$^*$\thanks{This work was partly supported by the National Natural Science Foundation of China under Grant No. 61450110086.},\\ Karl Henrik Johansson$^{**}$, Carla Seatzu$^*$
\\~\\
$^{*}$ Department of Electrical and Electronic Engineering, University of Cagliari, Italy \\
\{mehran.zareh,mauro.franceschelli,seatzu\}@diee.unica.it \\~\\
$^{**}$ School of Electrical Engineering, Royal Institute of Technology, Stockholm, Sweden \\ \{dimos,kallej\}@kth.se}
\maketitle

\begin{abstract} In this paper consensus in second-order multi-agent
systems with a non-periodic sampled-data exchange among agents is investigated. The sampling is random with bounded inter-sampling intervals. It is assumed that each agent has exact knowledge of its own state at any time instant. The considered local interaction rule is PD-type. Sufficient conditions for stability of the consensus protocol to a time-invariant value are derived based on LMIs. Such conditions only require the knowledge of the connectivity of the graph modeling the network topology. Numerical simulations are presented to corroborate the theoretical results.
\end{abstract}

\IEEEpeerreviewmaketitle

\section{Introduction}

Due to its broad spectrum of applications, in the past years, a
large attention has been devoted to the consensus problem in multi-agent systems
 (MAS) \cite{qin2011second,ren2005survey,yu2010some,Zareh_consensus2}.
Sensor networks \cite{yu2009distributed,olfati2005consensus}, automated highway systems \cite{ren2005survey},
mobile robotics \cite{khoo2009robust}, satellite alignment \cite{ren2007distributed} and several more,
are some of the potential areas in which a consensus problem is taken into account. Consensus is a state of a networked multi-agent system in which all the agents reach agreement on a common value by only sharing information locally, namely with their neighbors. Several algorithms, often called {\em consensus protocols}, have been proposed that lead a MAS to consensus. In particular, the coordination problem of mobile robots finds several applications in the manufacturing industry in the context of automated material handling. The consensus problem in the context of mobile robots consists in the design of local state update rules which allow the network of robots to rendezvous at some point in space or follow a leading robot exploiting only measurements of speeds and relative positions between neighboring robots. Robots are hereafter referred to as agents.

In MAS, heavy computational loads can interrupt the sampling period of a certain controller. A scheduled sampling period can be used to deal with this problem. In such a case robust stability analysis with respect to the changes in the sampling time is necessary. For interesting contributions in this area we address the reader to \cite{ackermann1985sampled,fridman2010refined,zutshi2012timed} and the references therein. We also mention the work by Fridman {\em et al.} \cite{fridman2004robust} who exploited an approach for time-delay systems and obtained the sufficient stability conditions based on the Lyapunov-Krasovskii functional method. Seuret \cite{seuret2012novel} and Fridman \cite{fridman2010refined} proposed methods with better upper bounds to the maximum allowed sampling. Shen {\em et al.} \cite{shen2012sampled} studied the sampled-data synchronization control problem for dynamical networks. Qin {\em et al.} \cite{qin2010sampled} and Ren and Cao \cite{ren2008convergence} studied the consensus problem for networks of double integrators with a constant sampling period. In the latter two papers, even though the authors use the sampled-data notation to introduce their novelty, they suppose that the communication and the local sensing occur simultaneously and this simplifies the problem into a discrete state consensus problem. Xiao and Chen \cite{xiao2012sampled} and Yu {\em et al.} \cite{yu2011second} studied second-order consensus in multi-agent dynamical systems with sampled \emph{position} data.

In this paper, we consider the case in which each agent has a perfect knowledge of its own state with almost no delay, i.e., it knows its own speed and position. Information exchanges between neighboring agents happen at discrete time intervals which are possibly non-periodic but strictly positive and bounded. The network dynamics can thus be modeled as a \emph{sampled-data system} (SDS), a class of systems extensively investigated in the literature. Using PD-like algorithm we guarantee that all the agents reach consensus. We recently proposed such a protocol in \cite{ETFA2014} where we provided a characterization of the convergence properties exploiting a Lyapunov-Krasovskii functional method. In particular in \cite{ETFA2014} we provided sufficient conditions for exponential stability of the consensus protocol to a time-invariant value under the assumption that the spectrum of the weighted adjacency matrix is known. In this paper we relax such assumption and provide sufficient conditions for consensus under the assumption that the only information on the network topology is its connectivity, i.e., the second largest eigenvalue of the weighted adjacency matrix. This is obviously a significant improvement with respect to \cite{ETFA2014}, not only because much less information on the network topology is needed, but also because, despite of \cite{ETFA2014}, the number of LMIs that have to be computed does not depend on the number of agents.

The paper is organized as follows. In Section~\ref{section:Prob_for} some notation and preliminaries are introduced. In Section~\ref{ProblemState} the consensus problem for second order multi-agent systems with non-periodic sampled-data exchange is formalized. In Section~\ref{main_res_1} the convergence properties of the proposed consensus protocol are characterized. In Section~\ref{sim_res} simulation results are presented to corroborate the theoretical analysis. Finally, in Section~\ref{conclusions} concluding remarks and directions for future research are discussed.

\section{Notation and Preliminaries}\label{section:Prob_for}

In this section we recall some basic notions on graph theory and introduce the notation used in the paper.

The topology of bidirectional communication channels
among the agents is represented by an undirected graph $\mathcal{G}=(\mathcal{V}, \mathcal{E})$ where $\mathcal{V}=\{1,\ldots,n\}$ is the set of nodes (agents) and $\mathcal{E}\subseteq\{\mathcal{V}\times \mathcal{V}\}$ is the set of edges. An edge $(i, j) \in  \mathcal{E}$ exists if there is a
communication channel between agent $i$ and $j$. Self loops
$(i, i)$ are not considered. The set of neighbors of agent $i$
is denoted by $\mathcal{N}_i  = \{j \ : \ (j, i) \in  \mathcal{E}; j = 1, \ldots, n\}$. Let
$\delta_i = |\mathcal{N}_i|$ be the degree of agent $i$ which represents the total
number of its neighbors.

The topology of graph $\mathcal{G}$ is encoded by the so-called {\em adjacency matrix}, an $n \times n$ matrix $A_d$ whose $(i,j)$-th entry is equal to $1$ if $(i,j)\in \mathcal{E}$, $0$ otherwise. Obviously in an undirected graph matrix $A_d$ is symmetric.

We denote $\Delta=\emph{diag}(\delta_{1}, \ldots,\delta_{n})$ the diagonal matrix whose non null entries are the degrees of the nodes. Moreover, matrix $W_d=\Delta^{-1} A_d$ is the {\em weighted adjacency matrix} associated with $\mathcal{G}$.
The following result has been proved in \cite{ETFA2014}.

\begin{lemma}\label{lemma_connected}
If a graph $\mathcal{G}$ is connected then the eigenvalues of the weighted adjacency matrix $W_d$, namely $\lambda_i, \ \, i=1,\ldots,n$, are all located in the interval $[-1, \ 1]$, and $\lambda_1=1$ is always a simple eigenvalue of $W_d$.
\end{lemma}

%


Finally, in the rest of this paper we denote with $*$ the symmetric elements of symmetric matrices.

\section{Problem Statement}\label{ProblemState}

Consider a second-order multi-agent system with an undirected communication topology. Consider the PD-type consensus protocol inspired by  \cite{cepeda2011exact} and \cite{Zareh_consensus}:
\begin{equation}\label{maineq_not_sampled}
\left\{
\begin{array}{lll}
 \dot{x}_i(t) & = & v_i(t),
\\
\dot{v}_i(t)& = & \dfrac{k_p}{\delta_i}\sum_{j \in \mathcal{N}_i}x_j(t)+\dfrac{k_d}{\delta_i}\sum_{j \in \mathcal{N}_i}v_j(t) \\
 & & -k_p x_i(t)-k_d v_i(t),
\end{array}
\right.
\end{equation}
where $n$ denotes the number of agents, $x_i(t)$ and $v_i(t)$ are the position and the velocity of agent $i$, and $\delta_i$ indicates its degree.

We suppose that the local information, i.e., the information that each agent receives from its own sensors, is measured instantaneously. This obviously makes sense when the sensor dynamics are fast enough. 

Moreover, we assume that the communication between the generic agent $i$ and its set of neighbors $\mathcal{N}_i$ occurs in stochastic sampling time instants $t_k$, $k=0,1,\ldots, \infty$, that satisfy the following conditions: $$0  < t_{k+1}-t_k\leq \bar{\tau}\in \mathbb{R}^+$$ and $$\lim\limits_{k\to \infty}t_k=\infty.$$

Under the above assumptions, equation \eqref{maineq_not_sampled} can be rewritten as:
\begin{equation}\label{eq_2_c}
\left\{
\begin{array}{lll}
 \dot{x}_i(t) & = & v_i(t),
\\
\dot{v}_i(t) & =  & \dfrac{k_p}{\delta_i}\sum_{j \in \mathcal{N}_i}x_j(t_k)+\dfrac{k_d}{\delta_i}\sum_{j \in \mathcal{N}_i}v_j(t_k) \\
 & & -k_p x_i(t)-k_d v_i(t)
\end{array}
\right.
\end{equation}
or, alternatively, doing some simple manipulations, as:
\begin{equation}\label{main}
\left[\begin{array}{c}
\dot{x}(t) \\
\dot{v}(t)
\end{array} \right]=(A\otimes I_n)\left[\begin{array}{c}
{x}(t) \\
{v}(t)
\end{array}\right]+(B\otimes W_d)\left[\begin{array}{c}
{x(t_k)} \\
{v(t_k)}
\end{array}\right]
\end{equation}
where $x=[x_1,x_2,\ldots, x_n]$, $v=[v_1,v_2,\ldots,v_n]$, $\Delta=\emph{Diag}\{\delta_1,\delta_2,\ldots, \delta_n\}$, $A_d$ is the adjacency matrix, $W_d$ is the weighted adjacency matrix, and matrices $A $ and $B$ are equal, respectively, to:
\begin{equation}\label{def:AB}
A=\left[\begin{array}{cc}
0 &1  \\
-k_p &-k_d
\end{array} \right], \qquad B=\left[\begin{array}{cc}
0 &0  \\
k_p &k_d
\end{array} \right].
\end{equation}

A MAS with an undirected communication topology and following equation~\eqref{maineq_not_sampled}, is said to converge to a \emph{consensus state} if $$\lim\limits_{t \to \infty }|x_i(t)-x_j(t)|=0$$ and $$\lim\limits_{t \to \infty }|v_i(t)-v_j(t)|=0.$$

In this paper, given the value of the maximum admissible difference $\bar{\tau}$ between any two consecutive sampling time instants, and a communication topology whose connectivity is known to be smaller than or equal to a given value $\bar \lambda$, we aim at finding conditions that guarantee consensus to a fixed point among agents that evolve according to equation~\eqref{main}.

%

\section{Convergence properties}\label{main_res_1}

In the following subsection we recall a state variable transformation, firstly introduced in \cite{ETFA2014}, to decouple the dynamics of modes associated with the eigenvalues of the weighted adjacency matrix. Then, the stability of such modes is analyzed in detail.

\subsection{Stability analysis}

Apply the following change of variables:
\begin{equation}
x(t)=T z(t)
\end{equation}
to eq.~\eqref{main}. Then, it holds:
\begin{equation}\label{transformed}
\begin{array}{lll}
(I_2 \otimes T)\left[\begin{array}{c}
\dot{z}(t) \\
\ddot{z}(t)
\end{array} \right] & = & (A\otimes  T)\left[\begin{array}{c}
{z}(t) \\
{\dot{z}}(t)
\end{array}\right] \\ & & +(B\otimes W_d T)\left[\begin{array}{c}
{z(t_k)} \\
{\dot{z}(t_k)}
\end{array}\right]
\end{array}
\end{equation}
and eq.~\eqref{main} can be rewritten as:
\begin{equation}\label{eq1_c}
\begin{array}{lll}
\left[\begin{array}{c}
\dot{z}(t) \\
\ddot{z}(t)
\end{array} \right] & = & (A\otimes I_n)\left[\begin{array}{c}
{z}(t) \\
{\dot{z}}(t)
\end{array}\right] \\ & & +(B\otimes T^{-1}W_d T)\left[\begin{array}{c}
{z(t_k)} \\
{\dot{z}(t_k)}
\end{array}\right].
\end{array}
\end{equation}
Since $W_d$ is a symmetrizable matrix, then it is also diagonalizable \cite{cepeda2011exact}, and the transformation matrix $T$ can be chosen such that $$\Lambda=T^{-1}W_d T=\emph{diag}(\lambda_1, \lambda_2, \ldots, \lambda_n)$$ where $$\lambda_1\geq \lambda_2\geq \ldots\geq \lambda_n$$ are the eigenvalues of the weighted adjacency matrix $W_d$.
As a result, eq.~\eqref{eq1_c} can be rewritten as:
$$\left[\begin{array}{c}
\dot{z}(t) \\
\ddot{z}(t)
\end{array} \right]=(A\otimes I_n )\left[\begin{array}{c}
{z}(t) \\
{\dot{z}}(t)
\end{array}\right]+(B\otimes \Lambda)\left[\begin{array}{c}
{z(t_k)} \\
{\dot{z}(t_k)}
\end{array}\right],$$
or alternatively, as
\begin{equation}
\left[\begin{array}{c}
\dot{z}_i(t) \\
\ddot{z}_i(t)
\end{array} \right]=A\left[\begin{array}{c}
{z}_i(t) \\
{\dot{z}}_i(t)
\end{array}\right]+\lambda_i B  \left[\begin{array}{c}
{z_i(t_k)} \\
{\dot{z}_i(t_k)}
\end{array}\right]
\end{equation}
where $i=1,\ldots,n$, and $z_i(t)$ is the $i$-th element of vector $z(t)$.

Now, if we define
\begin{equation}\label{def:mode}
y_i(t)=[z_i(t)\ \ \dot{z}_i(t)]^T
\end{equation}
the $i$-th {mode} of the system, we can say that its dynamics follows equation:
\begin{equation}\label{eq1}
\dot{y}_i(t)=Ay_i(t)+\lambda_i By_i(t_k).
\end{equation}

Moreover, assuming $\tau(t)=t-t_k$, the above equation can be rewritten as:
\begin{equation}\label{mode_dynamics}
\dot{y}_i(t)=Ay_i(t)+\lambda_i By_i(t-\tau(t)).
\end{equation}
The above SDS is a special case of a time varying delayed system where the delay $\tau(t)$ is upper bounded by $\bar{\tau}$, and its derivative is $\dot{\tau}(t)=1$, while the delay switches at times $t=t_k$, $k=0,1,\ldots, \infty$.

%
%


In the rest of this paper we assume that the graph $\mathcal{G}$ describing the communication topology is {\em connected}. By Lemma~\ref{lemma_connected} this implies that its largest eigenvalue is $\lambda_1=1$. We call \emph{unitary eigenvalue mode} (UEM) the mode associated with $\lambda_1=1$.


The following lemma, demonstrated in \cite{ETFA2014}, characterizes the dynamics of the UEM. In particular it shows that the UEM converges asymptotically to a vector whose first entry $z_1(t)$ is equal to a constant value and the second entry $\dot z_1(t)$ is null.

\begin{lemma}\label{lemma1}
Consider a system whose dynamics in the time interval $t\in [t_k,t_{k+1})$, $k=0,1,\ldots,\infty$, follows eq.~\eqref{eq1} with $i=1$ and $\lambda_i=1$. Assume $t_{k+1}-t_k>0$ for any $k=0,1,\ldots,\infty$. It holds

\begin{equation}
\lim_{k\rightarrow\infty} z_1(t_k)=\gamma, \quad \gamma \in \rea.
\end{equation}
\end{lemma}

We now provide the main contribution of this paper, i.e., we characterize the conditions on the design parameters $k_p, k_d, \bar \tau,\bar \lambda$ under which the modes $y_i(t)$, $i=2,\ldots,n$, defined in eq.~\eqref{def:mode} are asymptotically stable provided that $\lambda_i \leq \bar \lambda$ for all $i=2,\ldots,n$.

\begin{theorem}\label{new_theorem}
Consider the generic mode $y_i(t)$ defined in eq.~\eqref{def:mode} whose dynamics follows eq.~\eqref{mode_dynamics} where $\lambda_i$ is an uncertain parameter in $[-1,\bar \lambda]$, and obviously $\bar \lambda<1$.

If there exist positive definite matrices $P$ and $R$ and square matrices $Q_{1}$ and $Q_{2}$ such that the following inequalities hold:
\begin{equation}\label{LMI1}
M_{1}=\left[\begin{array}{cc}
 \begin{array}{c} Q_{1}^T(A-B)+ \\ (A-B)^TQ_{1} \end{array} & \begin{array}{c} P-Q_{1}^T+ \\ (A-B)^TQ_{2} \end{array}  \\~\\
 *& \begin{array}{c} -Q_{2}-Q_{2}^T  +\bar{\tau}R \end{array}
\end{array} \right]<0
\end{equation}
\begin{equation}\label{LMI2}
M_{2}=\left[\begin{array}{cc}
 \begin{array}{c} Q_{1}^T( A+\bar{\lambda} B)+ \\ ( A+\bar{\lambda} B)^TQ_{1} \end{array} &
 \begin{array}{c} P-Q_{1}^T+ \\ ( A+\bar{\lambda} B)^TQ_{2} \end{array}  \\~\\
 *& \begin{array}{c} -Q_{2}-Q_{2}^T+  \bar{\tau}R \end{array}
\end{array} \right]<0
\end{equation}
\begin{equation}\label{LMI3}
\begin{array}{l} M_3= \\~\\ \left[\begin{array}{ccc}
 \begin{array}{c} Q_{1}^T( A-  B)+\\ (  A- B)^TQ_{1} \end{array} & \begin{array}{c} P-Q_{1}^T+ \\ (  A-  B)^TQ_{2} \end{array} & \bar{\tau} Q_{1}^T B  \\~\\
 *&-Q_{2}-Q_{2}^T &\bar{\tau}  Q_{2}^T  B \\~\\
 * &* & -\bar{\tau}R
\end{array} \right] \\~\\ \qquad <0 \end{array}
\end{equation}
\begin{equation}\label{LMI4}
\begin{array}{l} M_4=\\~\\ \left[\begin{array}{ccc}
 \begin{array}{c} Q_{1}^T( A+\bar{\lambda} B)+\\ ( A+\bar{\lambda} B)^TQ_{1} \end{array} & \begin{array}{c} P-Q_{1}^T+ \\ ( A+\bar{\lambda} B)^TQ_{2}\end{array} &-\bar{\tau} {\bar \lambda} Q_{1}^T B  \\~\\
 *&  -Q_{2}-Q_{2}^T &-\bar{\tau} {\bar \lambda} Q_{2}^T B \\~\\
 * &* & -\bar{\tau}R
\end{array} \right] \\~\\ <0 \end{array}
\end{equation}
then the system with dynamics \eqref{mode_dynamics} is asymptotically stable.
\end{theorem}

{\em Proof:} Consider the Lyapunov function
\begin{equation}\label{def_V1}
\begin{array}{lll} V(t,y_i(t),y_i(t_k))  = &  y_i^T(t)P y_i(t) \\
 & + \left(\bar{\tau}-\tau(t)\right)\displaystyle \int_{t_k}^{t}\dot{y_i}^T(s)R\dot{y_i}(s)ds. \end{array}
\end{equation}

It holds:
\begin{equation}\label{v_dot}
\begin{array}{lll}
\dot{V}(t,y_i(t),y_i(t_k)) =  2 \dot y_i^T(t)P y_i(t)
 \\ \hspace{3cm} \displaystyle -\int_{t_k}^{t}\dot{y_i}^T(s)R\dot{y_i}(s)ds +\\  \qquad \left(\bar{\tau}-\tau(t)\right)  \big(\dot{y_i}^T(t)R\dot{y_i}(t)-\dot{y_i}^T(t_k)R\dot{y_i}(t_k)\big).
 \end{array}
\end{equation}

To provide an upper bound to \eqref{v_dot} we use Jensen integral inequality:
\begin{equation}\label{jensen}
\begin{array}{c}
\displaystyle \int\limits_{t_k}^{t}\dot{y}_i^T(s)R\dot{y}_i(s)ds \leq \int\limits_{t_k}^{t}\dot{y}_i^T(s)ds R\int\limits_{t_k}^{t}\dot{y}_i(s)ds.
\end{array}
\end{equation}
Define $\xi_i(t)=\displaystyle \frac{1}{\tau(t)}\int\limits_{t_k}^{t}\dot{y}_i(s)ds$.

We get:
\begin{equation}\label{jensen2}
\begin{array}{c}
\displaystyle \int\limits_{t_k}^{t}\dot{y}_i^T(s)R\dot{y}_i(s)ds \leq  \tau(t) \xi^T_i(t)R\xi_i(t)
\end{array}
\end{equation}
From the descriptor method  \cite{fridman2001descriptor} we know:
\begin{equation}
\begin{array}{r}
[y_i(t)\ \ \dot{y}_i(t)]\left[\begin{array}{c}
 Q_{1}\\
Q_{2}
\end{array} \right] \cdot \\ \quad \cdot(( A+\lambda_i B)y_i(t)-\tau(t)\xi_i(t)-\dot y_i(t)) =0 \end{array}
\end{equation}
Adding this to the right side of the inequality in \eqref{v_dot} and using the inequality \eqref{jensen2} we obtain:
\begin{equation*}
\dot{V}\leq \eta_i^T(t)\Psi(\tau(t),\lambda_i) \eta_i(t)-(\bar{\tau}-\tau(t))\dot{y_i}^T(t_k)R\dot{y_i}(t_k),
\end{equation*}
where $$\eta_=[y^T_i(t)\ \ \dot{y}^T_i(t) \ \ \xi^T_i(t)]^T$$ and:
\begin{equation}
\begin{array}{c}
\Psi(\tau(t),\lambda_i)= \\~\\ \left[\begin{array}{ccc}
\begin{array}{c} Q^T_{1}\Gamma_i+  \Gamma_i^T Q_{1} \end{array} & \begin{array}{c} P-Q^T_{1} \\
+\Gamma_i^T Q_{2} \end{array} &-\tau(t)\lambda_i Q^T_{1} {B}  \\~\\
* & \begin{array}{c} -Q_{2}-Q^T_{2}+\\ (\bar\tau-\tau(t))R \end{array} &-\tau(t)\lambda_i Q^T_{2}{B}  \\~\\
* & * & -\tau(t)R
\end{array} \right] \end{array}
\end{equation}
where $$\Gamma_i=({A}+\lambda_i {B}).$$

Notice that  $(\bar{\tau}-\tau(t))\dot{y_i}^T(t_k)R\dot{y_i}(t_k)$ is always positive. Thus:
\begin{equation}
\dot{V}\leq \eta_i^T(t)\Psi(\tau(t),\lambda_i) \eta_i(t),
\end{equation}
Hence to prove the stability one needs to prove that $\Psi(\tau(t),\lambda_i)$ is negative definite.

Now define the following matrices:
\begin{equation}\label{phi_0}
\begin{array}{c}
\Phi_{i,0}(\lambda_i)= \\~\\ \left[\begin{array}{cc}
 Q_{1}^T\Gamma_i +\Gamma_i^TQ_{1}  &  P-Q_{1}^T +\Gamma_i^TQ_{2}  \\~\\
 * &  Q_{2}-Q_{2}^T +\bar{\tau}R
\end{array} \right]. \end{array}
\end{equation}
and
\begin{equation}\label{phi_tau}
\begin{array}{c}
\Phi_{i,\bar{\tau}}= \\~\\ \left[\begin{array}{ccc}
 \begin{array}{c} Q_{1}^T \Gamma_i+ \Gamma_i^TQ_{1}\end{array} & \begin{array}{c} P-Q_{1}^T+ \Gamma_i^TQ_{2} \end{array} &-\bar{\tau}\lambda_i Q_{1}^T  B  \\~\\
 *& Q_{2}-Q_{2}^T &-\bar{\tau}\lambda_i Q_{2}^T  B \\~\\
 * &* & -\bar{\tau}R
\end{array} \right]\end{array}
\end{equation}
Define $$\eta^\prime_{i}(t)=[y^T_i(t)\;\; \dot{y}^T_i(t)]^T.$$ One can show that:
\begin{equation}
\begin{array}{lll}
 \eta^T_i(t)\Psi(\tau(t),\lambda_i)\eta_i(t) = \\~\\ \dfrac{\bar{\tau}-\tau(t)}{\bar\tau}\eta^{\prime^T}_{i}(t)\Phi_{i,0}\eta^{\prime}_{i}(t)+  \dfrac{\tau(t)}{\bar\tau}\eta^T_{i}(t)\Phi_{i,\bar{\tau}}\eta_{i}(t) = \\~\\ \dfrac{\bar{\tau}-\tau(t)}{\bar\tau}\eta^{\prime^T}_{i}(t)\Big(\dfrac{\bar{\lambda}-\lambda_i}{\bar\lambda+1}M_1+ \dfrac{\lambda_i}{\bar\lambda+1}M_2\Big)\eta^{\prime}_{i}(t)+ \\
 \qquad \dfrac{\tau(t)}{\bar\tau}\eta^T_{i}(t)\Big(\dfrac{\bar{\lambda}-\lambda_i}{\bar\lambda+1}M_3 +\dfrac{\lambda_i}{\bar\lambda+1}M_4\Big)\eta_{i}(t)
\end{array}
\end{equation}
Define $\mu_\tau=\dfrac{\bar{\tau}-\tau(t)}{\bar\tau}$ and $\mu_\lambda=\dfrac{\bar{\lambda}-\lambda_i}{\bar\lambda+1}$.

Then $\dfrac{\tau(t)}{\bar\tau}=1-\mu_\tau$, $\dfrac{\lambda_i}{\bar\lambda+1}=1-\mu_\lambda$ and
\begin{equation}
\begin{array}{lll}
 \eta^T_i(t)\Psi(\tau(t),\lambda_i)\eta_i(t)= \\~\\ \mu_\tau\eta^{\prime^T}_{i}(t)\Phi_{i,0}\eta^{\prime}_{i}(t)+(1-\mu_\tau)\eta^T_{i}(t)\Phi_{i,\bar{\tau}}\eta_{i}(t)= \\~\\ \mu_\tau\eta^{\prime^T}_{i}(t)\Big(\mu_\lambda M_1+(1-\mu_\lambda)M_2\Big)\eta^{\prime}_{i}(t)+\\
  (1-\mu_\tau)\eta^T_{i}(t)\Big(\mu_\lambda M_3+(1-\mu_\lambda)M_4\Big)\eta_{i}(t).
\end{array}
\end{equation}

Since $\mu_\tau\in [0,1]$  and $\mu_\lambda\in [0,1]$, coefficients $\mu_\tau$, $1-\mu_\tau$, $\mu_\lambda$, and $1-\mu_\lambda$ are positive. Moreover, by equations \eqref{LMI1} to \eqref{LMI4} it follows that $\Psi(\tau(t),\lambda_i)$ is negative definite and this proves the stability of the system. \hfill $\square$

\subsection{Consensus among agents}

We now prove the consensus of agents to a common position.

\begin{theorem}
Consider a MAS evolving according to equation \eqref{eq_2_c} where $\bar \tau$ is such that $0<t_{k+1}-t_k<\bar \tau<\infty$. Assume that the undirected connected graph $\mathcal{G}$ modeling the network topology is such that the second largest eigenvalue of its weighted adjacency matrix is smaller than or equal to $\bar \lambda$. If the LMIs defined in eq.~\eqref{LMI1} to \eqref{LMI4} are satisfied, then there exists a $\gamma \in \rea$ such that $x(t)$ asymptotically converges to $\gamma \vec{1}$ and $v(t)$ asymptotically converges to $\vec{0}$.
\end{theorem}

{\em Proof:} By Theorem~\ref{new_theorem}, if the LMIs in eq.~\eqref{LMI1} to \eqref{LMI4} hold, all modes except the UEM are asymptotically stable, i.e., $\lim\limits_{t\to \infty}y_i(t)=0$ and thus $ \lim\limits_{t\to \infty}z_i(t)=0$ for $i=2,\ldots,n$. Furthermore, by Lemma~\ref{lemma1}, there exists a positive constant $\gamma \in \rea$ such that $\lim\limits_{t\to \infty}z_1(t)=\gamma$.

Now, the first column of $T$ is the eigenvector corresponding to the unitary eigenvalue of $W_d$, therefore it is equal to $\vec 1=[1\ \ 1\ \ \ldots,\ \ 1 ]^T$. Thus, being $x(t)=T [z_1(t) \ \ 0 \ \ \ldots \ \ 0]^T$, it is trivial to show that when $t \to \infty$ it is $x_i(t)=x_j(t)$, for all $i,j=1,\ldots, n$.
The same calculations can be repeated for the velocities, thus proving that for $t \to \infty$, it is $v_i(t)=v_j(t)$, $i,j=1,\ldots, n $. \hfill $\square$

\section{Simulation results}\label{sim_res}

In this section we present the results of some numerical simulations that show the effectiveness of the proposed consensus protocol.
To this aim we consider a system with $8$ agents and assume $k_p=1$ and $k_d=1$.

\begin{figure}
\centering
\includegraphics[width=.9\linewidth]{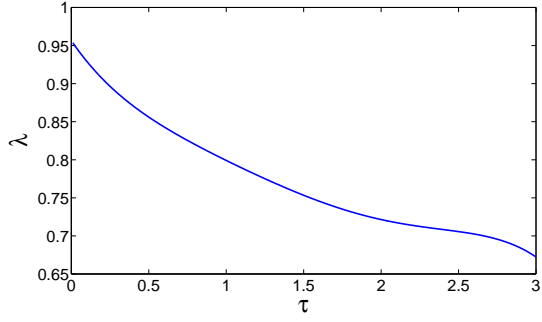}
\caption{The stability area in the $\bar\lambda-\bar\tau$ plane.}
\label{fig:tau-lambda}
\end{figure}

In Fig.~\ref{fig:tau-lambda} the area under the curve shows the stability region in the $\bar\lambda-\bar\tau$ plane. Such an area has been computed using the LMIs \eqref{LMI1} to \eqref{LMI4}.

We now consider a graph with adjacency matrix (randomly generated) equal to:
\begin{equation}
A_d=\left[\begin{array}{cccccccc}
 0 & 0 & 0 & 1 & 0 & 0 & 0 & 1 \\
 0 & 0 & 1 & 1 & 0 & 0 & 0 & 0 \\
 0 & 1 & 0 & 1 & 0 & 0 & 0 & 0 \\
 1 & 1 & 1 & 0 & 1 & 0 & 0 & 1 \\
 0 & 0 & 0 & 1 & 0 & 1 & 1 & 0 \\
 0 & 0 & 0 & 0 & 1 & 0 & 1 & 0 \\
 0 & 0 & 0 & 0 & 1 & 1 & 0 & 0 \\
 1 & 0 & 0 & 1 & 0 & 0 & 0 & 0
\end{array} \right].
\end{equation}

\begin{figure}
\centering
\includegraphics[width=.9\linewidth]{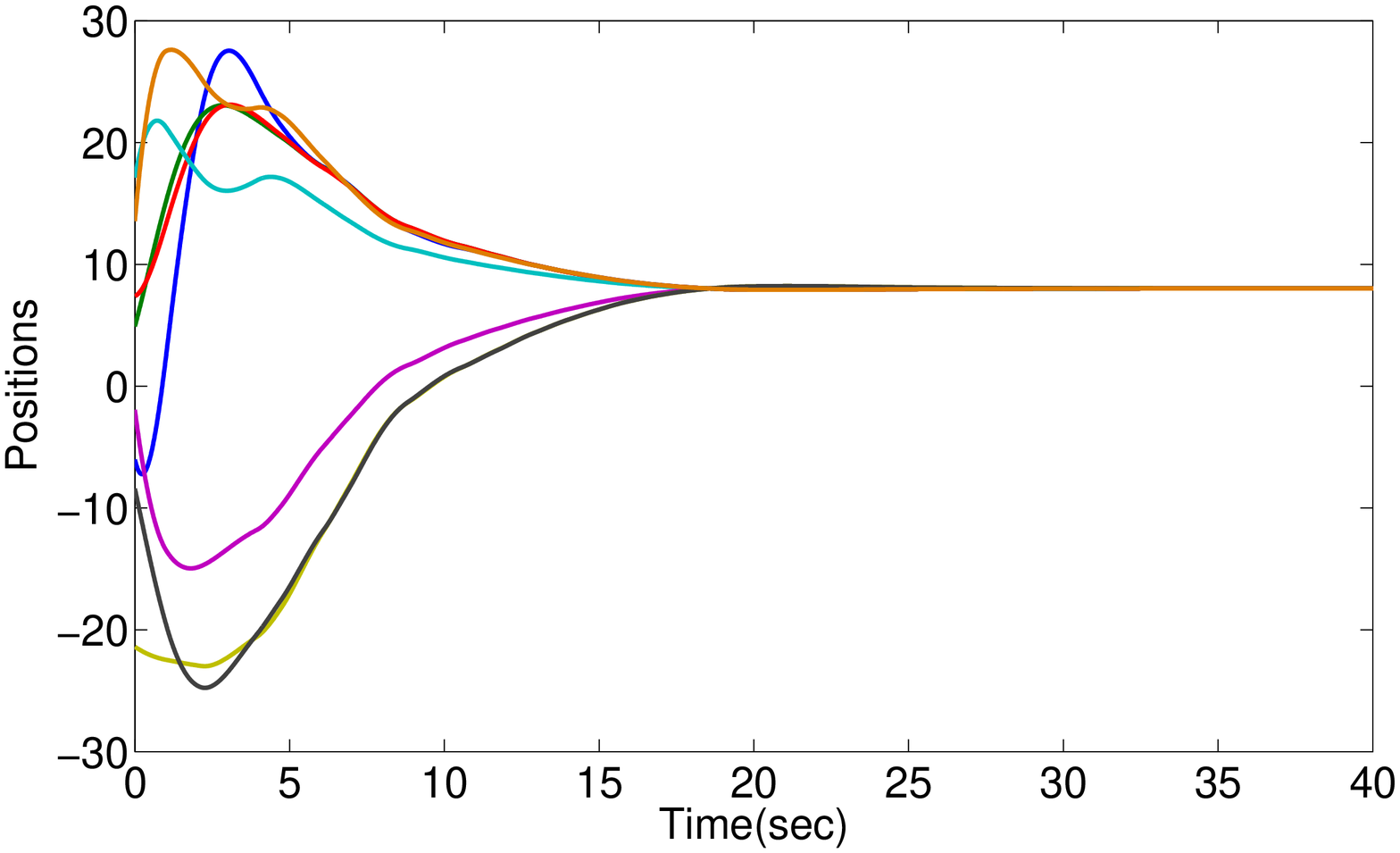}
\includegraphics[width=.9\linewidth]{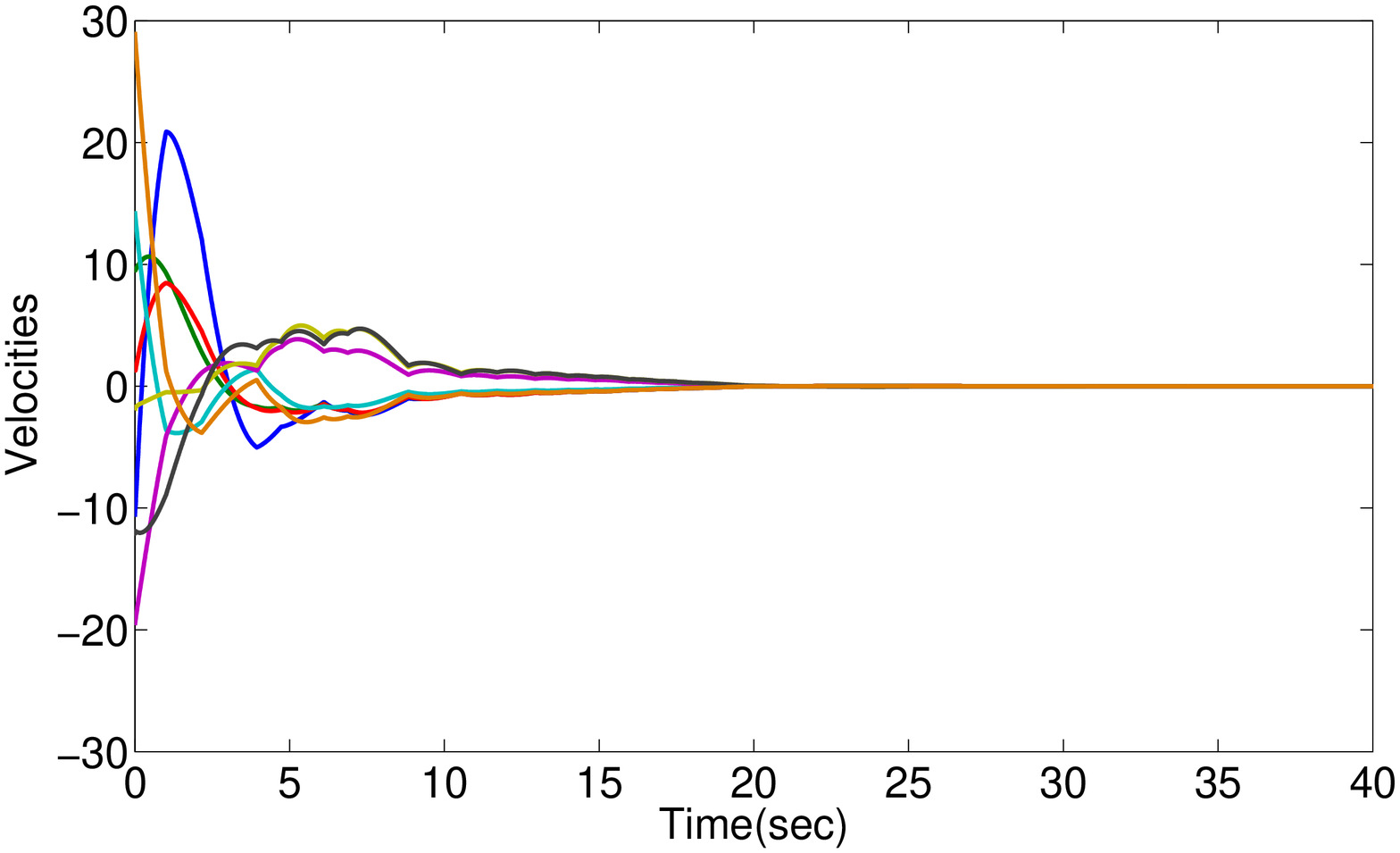}
\caption{Positions and velocities when the proposed protocol is implemented.}
\label{fig:8-agent-positions}
\end{figure}

\begin{figure}
\centering
\includegraphics[width=.9\linewidth]{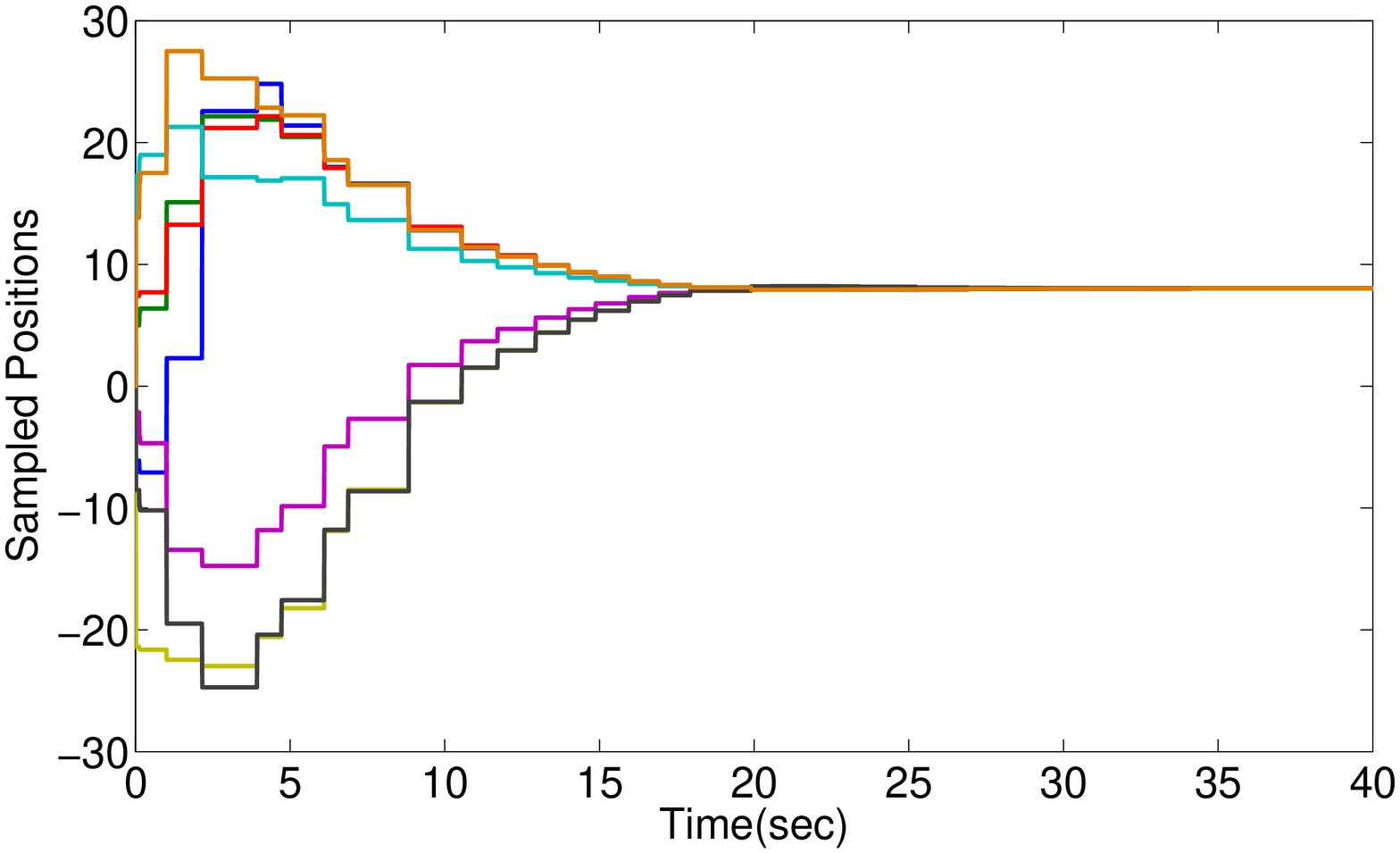}
\includegraphics[width=.9\linewidth]{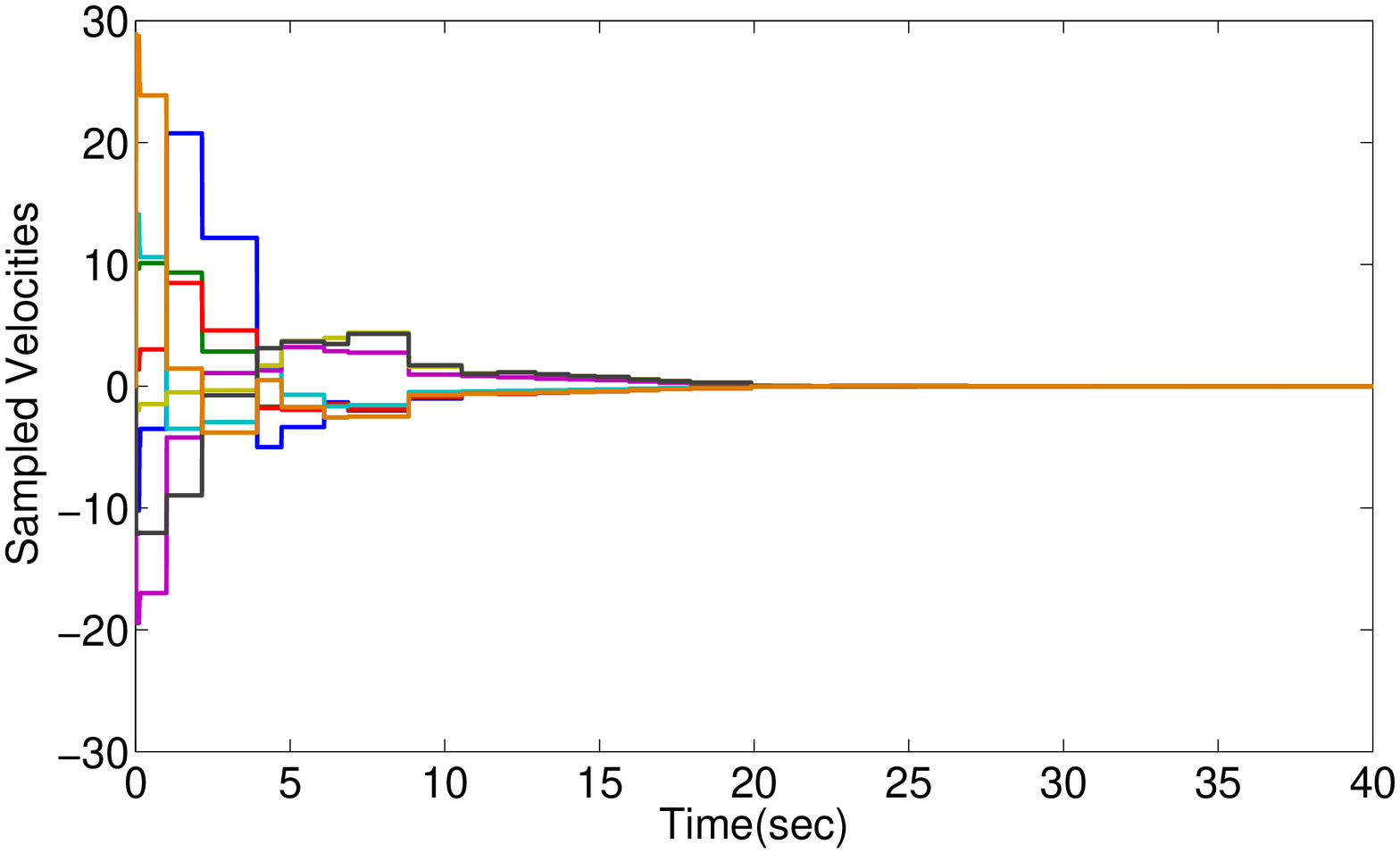}
\caption{Sampled positions and velocities aperiodically transmitted to neighbors by agents when the proposed protocol is implemented.}
\label{fig:8-agent-sampled velocities}
\end{figure}

Fig.~\ref{fig:8-agent-positions} shows the positions and velocities of the agents, while Fig.~\ref{fig:8-agent-sampled velocities} shows the sampled positions and velocities aperiodically transmitted to neighbors by each agent.

\section{Conclusions and future work}\label{conclusions}

In this paper we considered a PD-like consensus algorithm for a second-order multi-agent system where, at non-periodic sampling times, agents transmit to their neighbors information about their position and velocity, while each agent has a perfect knowledge of its own state at any time instant.
The main contribution consists in proving consensus to a common fixed point, based on LMIs verification, under the assumption that the network topology is not known and the only information is an upper bound on the connectivity.

Two are the main directions of our future research in this framework.
First, we want to compute analytically an upper bound on the value of the second largest eigenvalue of the weighted adjacency matrix that guarantees consensus, as a function of the other design parameters. Second, we plan to study the case where agents do not have a perfect knowledge of their own state.


\bibliographystyle{IEEEtran}

\bibliography{IEEEabrv,biblio_codit}

\begin{thebibliography}{10}
\providecommand{\url}[1]{#1}
\csname url@samestyle\endcsname
\providecommand{\newblock}{\relax}
\providecommand{\bibinfo}[2]{#2}
\providecommand{\BIBentrySTDinterwordspacing}{\spaceskip=0pt\relax}
\providecommand{\BIBentryALTinterwordstretchfactor}{4}
\providecommand{\BIBentryALTinterwordspacing}{\spaceskip=\fontdimen2\font plus
\BIBentryALTinterwordstretchfactor\fontdimen3\font minus
  \fontdimen4\font\relax}
\providecommand{\BIBforeignlanguage}[2]{{%
\expandafter\ifx\csname l@#1\endcsname\relax
\typeout{** WARNING: IEEEtran.bst: No hyphenation pattern has been}%
\typeout{** loaded for the language `#1'. Using the pattern for}%
\typeout{** the default language instead.}%
\else
\language=\csname l@#1\endcsname
\fi
#2}}
\providecommand{\BIBdecl}{\relax}
\BIBdecl

\bibitem{qin2011second}
J.~Qin, H.~Gao, and W.~X. Zheng, ``Second-order consensus for multi-agent
  systems with switching topology and communication delay,'' \emph{Systems \&
  Control Letters}, vol.~60, no.~6, pp. 390--397, 2011.

\bibitem{ren2005survey}
W.~Ren, R.~W. Beard, and E.~M. Atkins, ``A survey of consensus problems in
  multi-agent coordination,'' in \emph{2005 American Control Conference}, 2005.

\bibitem{yu2010some}
W.~Yu, G.~Chen, and M.~Cao, ``Some necessary and sufficient conditions for
  second-order consensus in multi-agent dynamical systems,'' \emph{Automatica},
  vol.~46, no.~6, pp. 1089--1095, 2010.

\bibitem{Zareh_consensus2}
M.~Zareh, C.~Seatzu, and M.~Franceschelli, ``Consensus on the average in
  arbitrary directed network topologies with time-delays,'' in \emph{4th IFAC
  Workshop on Distributed Estimation and Control in Networked Systems}, 2013.

\bibitem{yu2009distributed}
W.~Yu, G.~Chen, Z.~Wang, and W.~Yang, ``Distributed consensus filtering in
  sensor networks,'' \emph{IEEE Trans. on Systems, Man, and Cybernetics. Part
  B: Cybernetics}, vol.~39, no.~6, pp. 1568--1577, 2009.

\bibitem{olfati2005consensus}
R.~Olfati-Saber and J.~S. Shamma, ``Consensus filters for sensor networks and
  distributed sensor fusion,'' in \emph{44th IEEE Conf. on Decision and Control
  and 2005 European Control Conference}, 2005.

\bibitem{khoo2009robust}
S.~Khoo, L.~Xie, and Z.~Man, ``Robust finite-time consensus tracking algorithm
  for multirobot systems,'' \emph{Mechatronics, IEEE/ASME Transactions on},
  vol.~14, no.~2, pp. 219--228, 2009.

\bibitem{ren2007distributed}
W.~Ren, ``Distributed attitude alignment in spacecraft formation flying,''
  \emph{International Journal of Adaptive Control and Signal Processing},
  vol.~21, no. 2-3, pp. 95--113, 2007.

\bibitem{ackermann1985sampled}
J.~Ackermann, \emph{Sampled-data control systems: analysis and synthesis,
  robust system design}.\hskip 1em plus 0.5em minus 0.4em\relax Springer-Verlag
  New York, 1985.

\bibitem{fridman2010refined}
E.~Fridman, ``A refined input delay approach to sampled-data control,''
  \emph{Automatica}, vol.~46, no.~2, pp. 421--427, 2010.

\bibitem{zutshi2012timed}
A.~Zutshi, S.~Sankaranarayanan, and A.~Tiwari, ``Timed relational abstractions
  for sampled data control systems,'' in \emph{Computer Aided
  Verification}.\hskip 1em plus 0.5em minus 0.4em\relax Springer, 2012, pp.
  343--361.

\bibitem{fridman2004robust}
E.~Fridman, A.~Seuret, and J.-P. Richard, ``Robust sampled-data stabilization
  of linear systems: an input delay approach,'' \emph{Automatica}, vol.~40,
  no.~8, pp. 1441--1446, 2004.

\bibitem{seuret2012novel}
A.~Seuret, ``A novel stability analysis of linear systems under asynchronous
  samplings,'' \emph{Automatica}, vol.~48, no.~1, pp. 177--182, 2012.

\bibitem{shen2012sampled}
B.~Shen, Z.~Wang, and X.~Liu, ``Sampled-data synchronization control of
  dynamical networks with stochastic sampling,'' 2012.

\bibitem{qin2010sampled}
J.~Qin, W.~X. Zheng, and H.~Gao, ``Sampled-data consensus for multiple agents
  with discrete second-order dynamics,'' in \emph{49th IEEE Conf. on Decision
  and Control}, 2010.

\bibitem{ren2008convergence}
W.~Ren and Y.~Cao, ``Convergence of sampled-data consensus algorithms for
  double-integrator dynamics,'' in \emph{47th IEEE Conf. on Decision and
  Control}, 2008.

\bibitem{xiao2012sampled}
F.~Xiao and T.~Chen, ``Sampled-data consensus for multiple double integrators
  with arbitrary sampling,'' \emph{IEEE Trans. on Automatic Control}, vol.~57,
  no.~12, pp. 3230--3235, 2012.

\bibitem{yu2011second}
W.~Yu, W.~X. Zheng, G.~Chen, W.~Ren, and J.~Cao, ``Second-order consensus in
  multi-agent dynamical systems with sampled position data,''
  \emph{Automatica}, vol.~47, no.~7, pp. 1496--1503, 2011.

\bibitem{ETFA2014}
M.~Zareh, D.~V. Dimarogonas, M.~Franceschelli, K.~H. Johansson, and C.~Seatzu,
  ``Consensus in multi-agent systems with second-order dynamics and
  non-periodic sampled-data exchange,'' in \emph{19th IEEE Conf. on Emerging
  Technologies and Factory Automation}, Barcelona, Spain, 2014.

\bibitem{cepeda2011exact}
R.~Cepeda-Gomez and N.~Olgac, ``An exact method for the stability analysis of
  linear consensus protocols with time delay,'' \emph{IEEE Trans. on Automatic
  Control}, vol.~56, no.~7, pp. 1734--1740, 2011.

\bibitem{Zareh_consensus}
M.~Zareh, C.~Seatzu, and M.~Franceschelli, ``Consensus of second-order
  multi-agent systems with time delays and slow switching topology,'' in
  \emph{10th IEEE Int. Conf. on Networking, Sensing and Control}, 2013.

\bibitem{fridman2001descriptor}
E.~Fridman, ``A descriptor system approach to nonlinear singularly perturbed
  optimal control problem,'' \emph{Automatica}, vol.~37, no.~4, pp. 543--549,
  2001.

\end{thebibliography}

\end{document}